\documentclass{article}
\usepackage[affil-it]{authblk}
\usepackage{bm} %bold math type
\usepackage{subfigure}
\usepackage{amsthm}
\usepackage{amsmath}
\usepackage{graphicx}
\usepackage{amssymb}
\usepackage{geometry}
\usepackage{url}

\usepackage{algorithm2e}
\usepackage{natbib}
 \bibpunct[, ]{(}{)}{,}{a}{}{,}%
 \def\BIBand{and}%
\usepackage{bm}
\newtheorem{proposition} {Proposition}

\newtheorem{lemma} {Lemma}

\title{Robust Optimal Design of Two-Armed Trials with Side Information}

\author[1]{Qiong Zhang}
\author[2]{Amin Khademi}
\author[2]{Yongjia Song}

\affil[1]{ School of Mathematical and Statistical Sciences, Clemson University}
\affil[2]{Department of Industrial Engineering, Clemson University}
\date{}

\begin{document}

\maketitle

\begin{abstract}
Significant evidence has become available that emphasizes the importance of personalization in medicine. In fact, it has become a common belief that personalized medicine is the future of medicine. The core of personalized medicine is the ability to design clinical trials that investigate the role of patient covariates on treatment effects. In this work, we study the optimal design of two-armed clinical trials to maximize accuracy of statistical models where the interaction between patient covariates and treatment effect are incorporated to enable precision medication. Such a modeling extension leads to significant complexities for the produced optimization problems because they include optimization over design and covariates concurrently. We take a robust optimization approach and minimize (over design) the maximum (over population) variance of interaction effect between treatment and patient covariates. This results in a min-max bi-level mixed integer nonlinear programming problem, which is notably challenging to solve. To address this challenge, we introduce a surrogate model by approximating the objective function for which we propose two solution approaches. The first approach provides an exact solution based on reformulation and decomposition techniques. In the second approach, we provide a lower bound for the inner optimization problem and solve the outer optimization problem over the lower bound. We test our proposed algorithms with synthetic and real-world data sets and compare it with standard (re-)randomization methods. Our numerical analysis suggests that the lower bounding approach provides high-quality solutions across a variety of settings. 
\end{abstract}

%\KEYWORDS{deterministic inventory theory; infinite linear programming duality; 
%  existence of optimal policies; semi-Markov decision process; cyclic schedule}


\section{Introduction}
The average cost of bringing a new treatment to market has surpassed \$2.6 billion and expensive clinical trials are the major driver of such a high cost \citep{Tufts14}. In particular, the total cost of clinical trials can reach \$300-\$600 million for large global trials, and the costs usually increase with each phase of the trial \citep{giffin2010transforming}. Clinical trial costs depend on a variety of factors such as the number of participants, the number and locations of research facilities, the complexity of the trial protocol, and the reimbursement provided to investigators. 
In particular, the top three cost drivers of clinical trial expenditures are clinical procedure, administrative staff, and site monitoring costs \citep{sertkaya2016key}.
%For instance, from 1999 to 2005, the average length of a clinical trial has increased by 70\%; the average number of routine procedures increased by 65\%; and the average staff work burden increased by 67\% \citep{kaitin2008growing}. Despite their high complexity and costs, there are plenty of opportunities to improve the way clinical trials are conducted. 
Several different communities including statistics/biostatistics, public health sciences, economics,  machine learning, and operations research have studied different aspects of this complex procedure. 

Two-armed trials, i.e., trials with two treatment options, frequently appear in different phases of clinical studies. For instance, phase III clinical trials usually involve two treatments: one standard treatment or placebo and one candidate treatment. Phase III clinical trials are the most expensive ones and their improvement can significantly impact the efficiency of the procedure and statistical inferences \citep{giffin2010transforming}.
In addition, in the early stages of developing drugs, due to toxicity concerns, the decision makers usually start with two safe dosages (concentrations) of drug and run an experiment to test if the proposed drug has desirable efficacy. If successful, they increase the dosage and run a similar experiment with two dosages (one from previous experiment and one new increased dosage) until a satisfactory response is observed or the conclusion is that the drug is ineffective. Because in the early stages the efficacy of treatment and the magnitude of side effects are uncertain, the decision makers should be conservative in drawing conclusions \citep{le2009dose}. Motivated by these aforementioned settings, we investigate optimal designs for clinical trials with two treatment options. 
%In the process of developing a new treatment, dose-finding clinical trials play a critical role, where the goal is to identify the right treatment dosage. In fact, there is a trade-off that has to be carefully addressed: if the decision makers select an inadequate dose or doses that are too high, the current and/or next Phase III clinical trials will likely fail to receive approval from regulatory authorities \citep{bornkamp2007innovative}. As a result, the optimal design of dose-finding experiments is an essential phase and has received significant attention. Most of the works in this line of literature seek to derive optimal designs that minimize the asymptotic variance of a target dose under a specific dose-response relationship (see \cite{dette2008optimal} and the references therein for details). However, these studies model a population-level dose-response relationship and do not include individual-level analysis.

Specifically, we aim to incorporate patients' covariate information into the optimal design. This is motivated by the recent significant interest in personalized (precision) medicine. Personalized (precision) medicine seeks to maximize the quality of health care by providing individual-level health care for each patient and has recently gained prominence as the future of health care \citep{kosorok2019precision}. In fact, there is significant evidence that ignoring patient individualized information in prescribing medicine can impact the efficacy of treatment and potentially be harmful. For example, \cite{schork2015personalized} provided surprising statistics that the top ten highest-grossing drugs help between 1 in 25 and 1 in 4 of the patients. This number for statins can be as low as 1 in 50. Motivated by such evidence, governments and health care institutions have emphasized precision medicine and allocated significant resources for research and development in this area \citep{hayden2015california}. The majority of the literature investigates the optimal decision making of personalized treatment based on statistical analysis, see, e.g., \cite{shi2018maximin}. 

A key step in personalized medicine is the ability to design clinical trials that focus on an individual, not average, response to treatment \citep{schork2015personalized}. The key difference between precision medicine and population-level treatment in terms of their statistical analysis lies in the difference between their respective statistical models. This means that experiment designs originally developed for population-level treatment analysis may be inappropriate for precision medicine: it may deteriorate the accuracy and efficiency of their model estimation. It is critically important to investigate optimal design specifically for precision medicine, i.e., how to collect experimental data aiming to optimize the effectiveness of the subsequent statistical analysis for precision medication. To the best of our knowledge, this setting has not been addressed in the literature. To fill this gap, this paper extends a conventional approach for optimal design of clinical trials by incorporating patients' personalized covariate information, focusing on two-armed clinical trials. Formally, given a set of patients with covariate information, we study how to allocate them to different treatments in order to maximize the worst-case accuracy (over covariates) of statistical inferences about the treatment efficacy. Note that there are many other types of designs for clinical trials in the literature, such as adaptive, and Bayesian; for a survey of different types of designs for clinical trials, see~\cite{berry2006bayesian,press2009bandit,kotas2018bayesian} and references therein.

The theory of optimal experiment design started with the early development by~\cite{fisher1936design}. Classical optimal designs focus on reducing the variabilities of parameter estimation in a statistical model. Different types of optimal designs are often led by optimizing different utility functions of the variance-covariance matrix of the estimated parameters \citep{wu2011experiments, morris2015history}. For example, the D-optimal design corresponds to an optimal solution of minimizing the determinant of the generalized variance matrix of the parameter estimates for the underlying statistical model. As a result of the complex objective function employed in an optimal design problem, the corresponding optimization problem is usually challenging to solve. Off-the-shelf optimization solvers are usually incapable of providing exact optimal solutions of these optimization problems, see, e.g., \cite{singh2018approximation}.

Specialized solution methodologies must be developed to address this computational challenge. For instance, \cite{bertsimas2015power} have proposed a robust optimal design problem that minimizes the maximum discrepancy in mean and variance among different treatment groups. Their proposed designs yield a significant improvement over (re-)randomized designs in terms of statistical inference in the population level. Our paper incorporates precision medication in statistical modeling, and develops optimal designs to improve the accuracy of this task. Thus, the resulted structure of our  optimization formulations are significantly different from the one proposed in \cite{bertsimas2015power}. Specifically, in the context of two-arm clinical trials, their problem reduces to a single-level mixed integer linear program, which can be adequately handled by an off-the-shelf optimization solver. In contrast, our problem corresponds to a min-max bi-level mixed integer nonlinear program, for which we propose specialized algorithms. One other related work is by \cite{bhat2017near}, who studied optimal design of experiments with covariates for near-optimal A-B testing both in offline and online settings. In their offline setting, similar to our work, they studied an optimal design of experiments with a linear response model. The major difference is that we incorporate the interaction effects between patient informations and treatment allocation to enable precision decision making, while they do not. This simplification allows them to use a tractable approximation for the optimal design problem that minimizes the variance of the estimator. This is in stark contrast with our min-max bi-level mixed integer nonlinear programming formulations, which are computationally difficult to solve.

\textbf{Main contributions.} We summarize our main contributions as follows: 
\begin{itemize}
\item First, we formulate the optimal design problem for a two-armed clinical trial by incorporating patient covariate information into \emph{treatment effect}. Previous attempts in the literature to model the patient response with covariates consider the covariates in a linear form in response, but the interaction between treatment and patient covariates is not included. However, as mentioned above, treatment effect may be patient dependent. Our models incorporate information about covariates in treatment effects, for which the optimization of statistical accuracy is formulated as a min-max bi-level optimization problem from a robust optimization perspective: the decision maker seeks to minimize the worst case (over covariates) variance of the interaction between effect treatment and patient covariates. 

\item Second, we propose a surrogate model by approximating the variance of the estimator. The core of the approximation is adopting an asymptotic balance design in the Taylor expansion of the objective function. Despite this approximation, the optimization problem is still a min-max bi-level mixed integer nonlinear program, for which we propose two solution approaches to solve it. The first approach solves the surrogate model exactly and is based on a reformulation of the min-max problem using decomposition techniques. The second approach provides a lower bound for the inner maximization problem and the outer minimization is carried out over this lower bound. The appealing feature of the lower bounding approach is that it yields a single-level optimization problem, which scales well with the size of the problem including large clinical trials with hundreds of patient covariates.

\item Finally, we apply our algorithms on several sets of synthetic data and a case study with real data for patient covariates from a large clinical trial for Warfarin. Our numerical results show that our proposed algorithms outperform the standard randomization and re-randomization methods that are widely used in the optimal design literature in all tested settings. In particular, our results show that the proposed lower bounding approach performs robustly in terms of the corresponding objective values of both the surrogate model and the original model. This observation suggests that the lower bounding approach can be a fast and reliable option for optimal design of precision clinical trials.
\end{itemize}

\textbf{Paper organization.}  Section \ref{se:formulation} motivates and formulates the problem. Section \ref{se:solution} provides a surrogate model and introduces two solution approaches for the proposed optimal design problem. Section \ref{se:numerical} summarizes our numerical study and Section \ref{se:conclusion} concludes the paper. 

\section{Problem Formulation} \label{se:formulation}
Following classical assumptions in the literature~\citep{atkinson2015optimum,laber2016using,qian2011performance}, we consider a linear model 
to describe the treatment-response relationship in the presence of covariate information. In particular, let $x\in \{-1,1\}$ denote the two treatment levels, $\mathbf z=(1, z_1, \ldots, z_{p-1})^\top \in \mathcal Z \subset \mathbb{R}^{p}$ with $p>1$ denote the non-controllable patient covariates, and $y \in \mathbb{R}$ be a numerical response. The treatment-response relationship is then given by: 
\begin{equation}\label{eq:lm}
    y=\mathbf z^\top\pmb\alpha + x\mathbf z^\top\pmb\beta+\varepsilon,
\end{equation}
where $\pmb\alpha=(\alpha_0, \alpha_1, \ldots, \alpha_{p-1})^\top$ and $\pmb{\beta}=(\beta_0, \beta_1, \ldots, \beta_{p-1})^\top$ are the linear coefficients, and $\varepsilon$ models randomness in response and follows a normal distribution $N(0, \sigma^2)$. The purpose of personalized medicine is to recommend patient-specific treatment. 
To that end, the decision maker seeks to find the best (in terms of maximal response) treatment for each patient given its covariate $\mathbf z \in \mathcal Z$, which is defined by  
\begin{equation}\label{eq:target}
    x^\ast(\mathbf z) := \mathrm{argmax}_{x\in \{-1, +1\}} \left\{\mathbf z^\top \pmb{\alpha}+x\mathbf z^\top \pmb{\beta}\right\}=\mathrm{argmax}_{x\in \{-1, +1\}} \left\{x\mathbf z^\top \pmb{\beta}\right\}.
\end{equation}
Apparently, the optimal decision in \eqref{eq:target} can be viewed as 
a function of the individual's covariate information $\mathbf z$. 
We observe that the objective in \eqref{eq:target}
can be expressed by 
\[
x\mathbf z^\top \pmb{\beta}=x\beta_0+\sum^{p-1}_{i=1}xz_i\beta_i.
\]
Thus, the difference between the two treatment decisions for each individual depends
on the significance of the coefficient parameters associated with covariates
$xz_1, \ldots, xz_{p-1}$, i.e., the interaction between treatment $x$
and the patient covariates. If coefficients $\beta_1, \ldots, \beta_{p-1}$ are zeros in the model, we see that the personalized optimal decision $x^\ast(\mathbf z)$
is reduced to the population level optimal decision $x^\ast=\mathrm{argmax}_{x\in \{-1, +1\}} x\beta_0$, as studied in~\cite{bhat2017near}. 
%Therefore, accurate decision on the optimal treatment is essentially determined by 
%the accuracy of the predicted value of the interaction term $x\mathbf z^\top \pmb{\hat{\beta}}$ with estimated interaction coefficients $\pmb{\hat{\beta}}$. Since $\mathrm{var}(x\mathbf z^\top \pmb{\hat{\beta}})=\mathrm{var}(\mathbf z^\top \pmb{\hat{\beta}})$, the aim of optimal design under this context is to reduce the variance of $\mathbf z^\top \pmb{\hat{\beta}}$. 
Compared to the work of~\cite{bhat2017near}, model~\eqref{eq:lm} significantly improves the relevance to precision medicine. On the other hand, it also notably increases the complexity of the statistical analysis and computation, as the resulting estimators are multidimensional and include patient covariates. We next elaborate on these challenges.

\paragraph{Robust optimal design with covariates: a min-max bi-level optimization problem.} Suppose that $n$ patients are recruited for the clinical trial, the covariate information of each patient $i$ is given by $\mathbf h_i \in \mathcal Z$, and all patient covariate information is represented by an ${n\times p}$ matrix $H=(\mathbf h_1, \ldots, \mathbf h_n)^\top$, where $^\top$ denotes matrix transpose. Let $x_i\in \{-1, +1\}$ denote the treatment prescribed to patient $i$ and let $\mathbf x=(x_1, \ldots, x_n)^\top$ be the treatment allocation of $n$ patients. 
After the trial is finished and all the responses of patients are collected,  the estimated coefficients $\pmb\alpha$ and $\pmb\beta$ in \eqref{eq:lm} can be expressed by $\hat{\pmb{\alpha}}(\mathbf x, H)$ and $\hat{\pmb\beta}(\mathbf x, H)$, which are functions of the design $\mathbf x$ and patient covariate information $H$. Based on estimates of the model parameters, the decision maker is able to infer the best treatment for each patient type with covariate $\mathbf z \in \mathcal Z$. Let $\hat x(\mathbf z; H, \mathbf x)$ denote the suggested treatment to patients with covariates $\mathbf z$ where the trial has patient information $H$ and allocation prescribed is $\mathbf x$. A natural choice for $\hat x(\mathbf z; H, \mathbf x)$ is then given by:
\begin{align}\label{eq:target1}
\hat x(\mathbf z; H, \mathbf x) &:=  \mathrm{argmax}_{x\in \{-1,+1\}} \  \mathbf z^\top \hat{\pmb\alpha}(\mathbf x, H)+x\mathbf z^\top \hat{\pmb\beta}(\mathbf x, H) \nonumber \\
&=\mathrm{argmax}_{x\in \{-1,+1\}}x\mathbf z^\top \hat{\pmb\beta}(\mathbf x, H).
\end{align}
Recall that the treatment effect in model \eqref{eq:lm} is identified by $\mathbf z^\top{\pmb\beta}$, which can be estimated by $\mathbf z^\top \hat{\pmb\beta}(\mathbf x, H)$. In order to have a higher precision in statistical inference, it is natural to minimize 
the variance of  $\mathbf z^\top \hat{\pmb\beta}(\mathbf x, H)$ for each individual value of $\mathbf z$. 
According to our assumptions, $\varepsilon$ in \eqref{eq:lm} follows a normal distribution $\varepsilon\sim N(0,\sigma^2)$, thus $\mathbf z^\top \hat{\pmb\beta}(\mathbf x, H)$ also follows a normal distribution with 
mean $\mathbf z^\top {\pmb{\beta}}$ and variance $\mathbf z^\top \Sigma_\beta(\mathbf x, H) \mathbf z$, where $\Sigma_\beta(\mathbf x, H)$ is the variance-covariance matrix of $\hat{\pmb\beta}(\mathbf x, H)$. From a robust optimization perspective, we aim to minimize the worst-case (maximum) variance of the estimated interaction effect $\mathbf z^\top \hat{\pmb\beta}(\mathbf x, H)$ among all patient covariates $\mathbf z \in \mathcal Z$, which yields the following optimization problem:
\begin{equation}\label{eq:var}
    \mathrm{min}_{\mathbf x\in \{-1, +1\}^n}\mathrm{max}_{\mathbf z\in \mathcal Z}\,\,\,\mathbf z^\top \Sigma_\beta(\mathbf x, H) \mathbf z.
\end{equation}

\paragraph{Challenges for solving the min-max bi-level optimization~\eqref{eq:var}.}
We next characterize the variance-covariance matrix $\Sigma_\beta(\mathbf x, H)$ in the objective function of~\eqref{eq:var}, and point out the challenges in solving this optimal design problem. Notice that, the dimension of covariates in \eqref{eq:lm} is $2p$ by including the main effect $\mathbf z$ and the interaction $x\mathbf z$. After stacking all the covariates from all $n$ patients, we denote the $n\times 2p$ covariates matrix by
\[
X=
\begin{bmatrix}
   H & D_{\mathbf x}H
\end{bmatrix},
\]
where $D_{\mathbf x}=\mathrm{diag}(x_1, \ldots, x_n)$ is a diagonal $n \times n$ matrix, and $H$ and $D_{\mathbf x}H$ are the matrices of patient covariates and the matrices that characterize interactions between treatment allocation and patient covariates, respectively. Thus, the variance matrix of  the estimated parameters $(\pmb \alpha^\top, \pmb\beta^\top)^\top$ in \eqref{eq:lm} can be expressed by
\[
\sigma^2(X^\top X)^{-1}=\sigma^2
\begin{bmatrix}
   H^\top H & H^\top D_{\mathbf x} H \\
   H^\top D_{\mathbf x} H & H^\top H
\end{bmatrix}^{-1}.\]
By taking the inversion of the above block matrix, the variance of the estimator for $\pmb\beta$ corresponds to the second diagonal block entry, which is given by:
\begin{equation} \label{eq:sigma}
\Sigma_\beta(\mathbf x, H)=\sigma^2\left(H^\top H-H^\top D_{\mathbf x} H (H^\top H)^{-1}H^\top D_{\mathbf x} H\right)^{-1}.
\end{equation}
Plugging equation \eqref{eq:sigma} into optimization problem \eqref{eq:var} results in a min-max bi-level non-convex mixed integer nonlinear program, which is notoriously difficult to solve. Furthermore, the covariate matrix in~\eqref{eq:sigma} makes the optimization problem challenging to handle directly due to the matrix inverse. The next section introduces our proposed approaches to address the computational challenges in solving~\eqref{eq:var}.

\section{Solution Approaches} \label{se:solution}
Our proposed solution approaches are based on an approximation to optimization problem~\eqref{eq:var} using a surrogate objective function. We describe this surrogate objective function in Section~\ref{sse:surrogate}. The construction of the surrogate model is based on a natural asymptotic result on the number of patients allocated to each treatment, i.e., as the number of patients increases, the optimal design converges to a balanced design, which is the gold standard in the literature, see, e.g.,~\cite{kallus2018optimal}. After applying this surrogate function, the problem remains a min-max bi-level non-convex mixed integer nonlinear program, and we subsequently describe two solution approaches. In Section \ref{sse:exact}, we provide an exact algorithm to solve this surrogate model based on decomposition and reformulation methods. In Section \ref{sse:LB}, we derive a lower bound for the inner optimization problem, which allows the min-max bi-level optimization problem to be further approximated by a single-level optimization problem by replacing the inner optimization problem with this lower bound. 

\subsection{A Surrogate Optimization Model} \label{sse:surrogate}
In this section, we approximate $\Sigma_\beta(\mathbf x, H)$ based on some natural assumptions in our problem context and then use this approximation to construct a surrogate optimization problem for~\eqref{eq:var}. Let $n_+$ and $n_-$ denote the number of patients that are allocated to treatments $+1$ and $-1$, respectively. In addition, let $\mathbb{P}_{\mathbf z}$ denote the probability measure defined over the covariate space, and $\mathbb{E}_{\mathbf z}$ denote the expectation with respect to the said probability measure. The next lemma is crucial is our construction, which ensures that under an asymptotically balanced design, the following matrix behaves asymptotically as $\frac{1}{n} I_p$, where $I_p$ is a $p \times p$ identity matrix.
\begin{lemma} \label{lm:balanced}
Let $\mathbb{E}_{\mathbf z} (z_l z_k)=\gamma_{lk}$ and assume that $|\gamma_{ll}| \geq \gamma > 0 , \ \forall l \in \{1,2,\ldots,p\}$ and $\frac{n_+}{n}-\frac{n_-}{n}=O(n^{-1})$, then
\[
(H^\top H)^{-1}H^\top D_{\mathbf x} H=O_p(n^{-1})I_p.
\]
\end{lemma}
\proof{Proof}
By law of large numbers, we have
\[
 n^{-1}\sum^n_{i=1}z_{il} z_{ik}=\gamma_{lk}+o_p(1).
\]
Let $\Gamma$ be a $(p+1)\times (p+1)$ matrix with $ij$-th entry $\gamma_{ij}$
As a result,
\[
H^\top H/n=\frac{1}{n}\sum^n_{i=1}\mathbf z_i \mathbf z^\top_i=\Gamma+o_p(1).
\]

\[
H^\top D_{\mathbf x} H/n=\frac{1}{n}\sum^n_{i=1}x_i\mathbf z_i \mathbf z^\top_i=\frac{1}{n}\sum_{\{i: x_i=1\}}\mathbf z_i \mathbf z^\top_i-\frac{1}{n}\sum_{\{i: x_i=-1\}}\mathbf z_i \mathbf z^\top_i
\]
\[
=\frac{n_+}{n}n^{-1}_+\sum_{\{i: x_i=1\}}\mathbf z_i \mathbf z^\top_i-\frac{n_-}{n}n^{-1}_-\sum_{\{i: x_i=-1\}}\mathbf z_i \mathbf z^\top_i
\]
\[
=(\frac{n_+}{n}-\frac{n_-}{n})(\Gamma+o_p(1))
\]
Consider  that $\frac{n_+}{n}-\frac{n_-}{n}=O(n^{-1})$, we have 
\[
(H^\top H)^{-1}H^\top D_{\mathbf x} H=O_p(n^{-1})I. \qquad \blacksquare
\]

\endproof

%Two observations are in order. 
%First, the assumption $\mathbb{E}_{\mathbf z} (z_l z_k)=\gamma_{lk}$ ensures that covariates are not independent ({\color{red} I think we do not need this assumption}). Second, 

The assumption $\frac{n_+}{n}-\frac{n_-}{n}=O(n^{-1})$ formalizes asymptotically balanced design concept, which is considered as a gold standard in optimal design of experiments. We provide numerical evidence regarding the quality of the above approximation in Appendix \ref{sec:compare}.

The covariate matrix in \eqref{eq:sigma} is difficult to manage because it does not give an explicit formula on $\mathbf x$, as a result of the matrix inverse. To simplify this expression, we apply the Taylor expansion of \eqref{eq:sigma}, and use results provided in Lemma~\ref{lm:balanced}.

\begin{proposition} \label{prop:approx}
Under the assumptions of Lemma \ref{lm:balanced} we have
\begin{equation}\label{eq:approx}
\Sigma_\beta(\mathbf x, H)
=\sigma^2 \left( I+ (H^\top H)^{-1}H^\top D_{\mathbf x} H (H^\top H)^{-1}H^\top D_{\mathbf x} H\right) (H^\top H)^{-1} + O(n^{-2}) A(\mathbf x, H),
\end{equation}
where $A(\mathbf x, H)$ is the remainder matrix for coefficients with $n^{-2}$ and polynomials of higher degrees in the Taylor's expansion.
\end{proposition}  
\proof{Proof}
Recall that 
\begin{align*}
\Sigma_\beta(\mathbf x, H)&=\sigma^2\left(H^\top H-H^\top D_{\mathbf x} H (H^\top H)^{-1}H^\top D_{\mathbf x} H\right)^{-1}\\
&= \sigma^2 \left ( (H^\top H) (I - (H^\top H)^{-1}H^\top D_{\mathbf x} H (H^\top H)^{-1}H^\top D_{\mathbf x} H)\right)^{-1}\\
&=  \sigma^2 \left(I - (H^\top H)^{-1}H^\top D_{\mathbf x} H (H^\top H)^{-1}H^\top D_{\mathbf x} H\right)^{-1} (H^\top H)^{-1}\\
&=  \sigma^2 \left( (I+(H^\top H)^{-1}H^\top D_{\mathbf x} H) (I-(H^\top H)^{-1}H^\top D_{\mathbf x}H) \right)^{-1} (H^\top H)^{-1} \\
&= \sigma^2 (I-(H^\top H)^{-1}H^\top D_{\mathbf x}H)^{-1} (I+(H^\top H)^{-1}H^\top D_{\mathbf x} H)^{-1} (H^\top H)^{-1} \\
&= \sigma^2 \left(I+(H^\top H)^{-1}H^\top D_{\mathbf x}+\sum_{k=2}^{\infty} ((H^\top H)^{-1}H^\top D_{\mathbf x}H)^k\right). 
\end{align*}
Therefore, 
\begin{align*}
&\left(I-(H^\top H)^{-1}H^\top D_{\mathbf x}+ \sum_{k=2}^{\infty} (-1)^k((H^\top H)^{-1}H^\top D_{\mathbf x}H)^k\right)(H^\top H)^{-1}\\
&= \sigma^2 \left( (I+(H^\top H)^{-1}H^\top D_{\mathbf x}+ O(n^{-2})I) (I-(H^\top H) ^{-1}H^\top D_{\mathbf x}+ O(n^{-2})I)\right) (H^\top H)^{-1}\\
&=\sigma^2 \left( I+ (H^\top H)^{-1}H^\top D_{\mathbf x} H (H^\top H)^{-1}H^\top D_{\mathbf x} \right) (H^\top H)^{-1} + O(n^{-2}) A(\mathbf x, H). \qquad \blacksquare
\end{align*}
\endproof

Proposition \ref{prop:approx} paves the way to construct the surrogate model. In fact, it is natural to replace $\Sigma_\beta(\mathbf x, H)$ by the first term of~\eqref{eq:approx}. To streamline notation, let 
\begin{equation}
\Psi (\mathbf x , H)=  (H^\top H)^{-1}H^\top D_{\mathbf x} H (H^\top H)^{-1}H^\top D_{\mathbf x} H(H^\top H)^{-1}.
\end{equation}

Then, a surrogate model for optimization problem~\eqref{eq:var} can be formulated as:
\begin{subequations}\label{eq:surrogate}
\begin{align}
    \mathrm{min}_{\mathbf x} &\mathrm{max}_{\mathbf z\in \mathcal Z}\,\,\,\mathbf z^\top ((H^\top H)^{-1}+\Psi (\mathbf x , H)) \mathbf z \label{eq:surrogate-1} \\
         \text{s.t. } &-1 \leq \sum^n_{i=1} x_i\leq 1, \label{eq:surrogate-2} \\
         &\mathbf x\in \{-1, +1\}^n. 
\end{align}
\end{subequations}

Note that compared to the original problem~\eqref{eq:var}, we add an additional constraint~\eqref{eq:surrogate-2} to the outer optimization problem to ensure a balanced design. Particularly, if $n$ is an even number, this constraint becomes the exact balancing constraint, i.e., $\sum^n_{i=1} x_i=0$. If $n$ is an odd number, exact balancing over $\mathbf x$ is impossible, thus we relax the constraint to be $-1 \leq \sum^n_{i=1} x_i\leq 1$. Before proceeding, the following lemma shows that the objective function of~\eqref{eq:surrogate} can be rewritten as a quadratic function of $\mathbf x$ given a fixed $\mathbf z$.
\begin{lemma}
The following equality holds
\begin{equation}\label{eq:reform-quadratic-x}
\mathbf z ^ \top  \Psi (\mathbf x , H) \mathbf z = \mathbf x ^\top \Upsilon (\mathbf z, H) \mathbf x,
\end{equation}
where 
\[
 \Upsilon (\mathbf z, H)= (H(H^\top H)^{-1}H^\top) \circ (H(H^\top H)^{-1}\mathbf z (\mathbf{z})^\top
(H^\top H)^{-1}H^\top),
\]
and $\circ$ denotes the matrix elementary-wise product. In addition, matrix $\Upsilon (\mathbf z, H)$ is positive semi-definite (PSD) for any $\mathbf z$.
\end{lemma}
\proof{Proof}
Let $\mathrm{tr}(\cdot)$ denote the trace of a matrix. The result follows by 
\[
(\mathbf{z})^\top
(H^\top H)^{-1}H^\top D_{\mathbf x} H (H^\top H)^{-1}H^\top D_{\mathbf x} H(H^\top H)^{-1}\mathbf z
\]
\[
=\mathrm{tr}\left\{(\mathbf{z})^\top
(H^\top H)^{-1}H^\top D_{\mathbf x} H (H^\top H)^{-1}H^\top D_{\mathbf x} H(H^\top H)^{-1}\mathbf z\right\}
\]
\[
=\mathrm{tr}\left\{ D_{\mathbf x} H (H^\top H)^{-1}H^\top D_{\mathbf x} H(H^\top H)^{-1}\mathbf z(\mathbf{z})^\top
(H^\top H)^{-1}H^\top\right\}
\]
\[
=\mathbf x^\top \left\{H (H^\top H)^{-1}H^\top \circ H(H^\top H)^{-1}\mathbf z(\mathbf{z})^\top
(H^\top H)^{-1}H^\top\right\} \mathbf x. \qquad \blacksquare
\]
\endproof

Although problem~\eqref{eq:surrogate} is still a min-max bi-level mixed integer nonlinear program, its objective function is much easier to handle: given a fixed $\mathbf x$, the objective function is a convex quadratic function of $\mathbf z$ (see \eqref{eq:surrogate-1}), and given a fixed $\mathbf z$, the objective function is a convex quadratic function of $\mathbf x$ (see \eqref{eq:reform-quadratic-x}). We next develop exact and heuristic approaches to solve it.

\subsection{An Exact Algorithm for Solving the Surrogate Model} \label{sse:exact}
In this section, we develop an exact algorithm to solve formulation~\eqref{eq:surrogate} by applying a cutting plane procedure on a reformulation of formulation~\eqref{eq:surrogate}. This reformulation is motivated by the fact that the objective function of problem~\eqref{eq:surrogate} is a convex quadratic function of $\mathbf x$ given a fixed $\mathbf z$, and a convex quadratic function of $\mathbf z$ given a fixed $\mathbf x$. Before proceeding, we make an assumption on the patient covariate space to facilitate the derivation of the algorithm. In particular, we assume that  $z_i \in \{-1,1\}$ for all $i=1,2,\ldots, p-1$, i.e., each covariate can be represented by a binary variable. This assumption results in $\mathcal Z = 1 \times \{-1,1\}^{p-1}$, where the first ``1'' indicates that the first covariate is set to be one (as the 
intercept). Note, however, that this assumption is made without loss of generality, since it is well-known that a linear model with discrete and bounded covariates can be transformed into an equivalent linear model in which all the covariates are binary. Also, in the context of clinical trials the covariates are usually categorical such as age group, gender, and health category, which are discrete and bounded. Therefore, one can easily construct a linear model with binary covariates. We denote $\mathcal Z= \{Z_1,Z_2,\ldots, Z_{2^{p-1}}\}$.

First, the surrogate model~\eqref{eq:surrogate} can be reformulated as:
\begin{align} \label{eq:reformulation}
    \min \ & \theta \\
    \text{s.t. } & \theta \geq \mathbf z^\top(H^\top H)^{-1}\mathbf z + \mathbf x^\top  \Upsilon (\mathbf z, H) \mathbf x, \quad \forall \mathbf{z}\in \mathcal Z\nonumber \\
    &-1 \leq \sum^n_{i=1} x_i\leq 1,\nonumber \\
   & \mathbf x \in \{-1,1\}^n. \nonumber
\end{align}
Note that the above formulation is a convex integer quadratic program, and we propose a cutting-plane based exact solution approach. In particular, let $\mathcal Z_m \subset \mathcal Z$, and define the following so-called master problem that gives a relaxation of~\eqref{eq:reformulation}:
\begin{align}\label{eq:reformulation-master}
    \min \ & \theta \\
     \text{s.t. } & \theta \geq \mathbf z^\top(H^\top H)^{-1}\mathbf z + \mathbf x^\top  \Upsilon (\mathbf z, H) \mathbf x, \quad \forall \mathbf z\in \mathcal Z_m\nonumber \\
     &-1 \leq \sum^n_{i=1} x_i\leq 1,\nonumber \\
   & \mathbf x \in \{-1,1\}^n. \nonumber
\end{align}
Let $\mathbf x_m$ and $\theta_m$ be an optimal solution to the master problem~\eqref{eq:reformulation-master}. If $\mathbf x_m$ and $\theta_m$ satisfy all of the constraints in formulation \eqref{eq:reformulation}, then, $\mathbf x_m$ and $\theta_m$ are optimal to~\eqref{eq:reformulation}. Otherwise, we should add elements in $ \mathcal Z \setminus \mathcal Z_m$ to $\mathcal Z_m$ and re-solve the master problem~\eqref{eq:reformulation-master} to obtain a tighter relaxation. To find these elements (if any), we solve the following subproblem: 
\begin{align}\label{eq:subproblem}
\delta_m= \max_{\mathbf z \in \mathcal Z}\,\,\,\mathbf  z^\top\left((H^\top H)^{-1}+ \Psi (\mathbf x_m, H)\right) \mathbf z.
\end{align}
If $\theta_m\ge \delta_m$, the optimal solution is $\mathbf x_m$; otherwise, we add an optimal solution of \eqref{eq:subproblem} to $\mathcal Z_m$ and continue the procedure by re-solving the master problem~\eqref{eq:reformulation-master}. Because $\mathcal Z$ is finite, this procedure converges to an optimal solution of \eqref{eq:reformulation} in a finite number of steps. In our implementation of the algorithm though, we use the stopping criteria of $\theta_m \ge \delta_m - \epsilon$ for some pre-specified threshold $\epsilon > 0$. Observe that the subproblem \eqref{eq:subproblem} is a nonconvex quadratic integer program, which is difficult to solve in general. However, recall that each $z_i \in \{-1,1\}$, then one can apply McCormick reformulation of bilinear and quadratic terms and the resulting reformulation will be a mixed integer linear program, which can be solved by an off-the-shelf optimization solver.

\subsection{A Lower Bounding Approximation to the Inner Maximization Problem of~\eqref{eq:surrogate}} \label{sse:LB}
In this section, we propose a lower bound for the inner maximization problem of~\eqref{eq:surrogate}, and we propose a heuristic approach for solving the surrogate model~\eqref{eq:surrogate} by solving a single-level optimization problem, which is obtained by replacing the inner maximization problem with this lower bound. To that end, we assume that  $\mathcal Z=\{\mathbf h_1, \ldots, \mathbf h_n\}$, i.e., the collection of covariate vectors of patients coincide with the space of all possible covariates of interest. This assumption, though, is not restrictive because when decision makers set up a clinical trial to investigate the efficacy of a drug for a specific set of covariates, they recruit patients with these covariates. More importantly, although the validity of the lower bounding technique relies on this assumption, the solutions produced by following this heuristic approach can be used even for settings where the assumption is not satisfied. Our numerical results show that the solutions provided by this approach are competitive with those derived by the exact algorithm presented in Section \ref{sse:exact} for the surrogate model, and it outperforms the exact algorithm in terms of the original objective~\eqref{eq:var} when the number of covariates is large: see Section \ref{sec:syn}.

\begin{proposition}
For any given $\mathbf x$, the inner maximization problem of~\eqref{eq:surrogate} is lower bounded by
\[
\frac{p}{n}+\frac{1}{n} \mathbf x^\top\left[(H(H^\top H)^{-1}H^\top)\circ (H(H^\top H)^{-1}H^\top)\right]\mathbf x.
\] 
\end{proposition}
\proof{Proof}
By letting $\mathcal Z=\{\mathbf h_1, \ldots, \mathbf h_n\}$, we observe that  $\sum^n_{i=1}\mathbf h_i\mathbf h^\top_i=H^\top H$. Let $\mathrm{tr}(\cdot)$ denote the trace of a matrix, then for a given $\mathbf x$, the inner problem in \eqref{eq:surrogate}: 
%\[
%p=\mathrm{tr}\left((H^\top H)^{-1}(H^\top H)\right)=\sum^n_{i=1}\mathrm{tr}\left((H^\top H)^{-1}\mathbf h_i \mathbf h^\top_i\right)
%\]
%\[\leq n 
%\mathrm{max}_{\mathbf z\in\mathcal Z}\mathrm{tr}\left((H^\top H)^{-1}\mathbf z\mathbf z^\top\right)=n\mathrm{max}_{\mathbf z\in\mathcal Z}\mathbf z^\top(H^\top H)^{-1}\mathbf z,
%\]
%where $\mathrm{tr}(\cdot)$ denotes the trace of a matrix. Thus, 
\begin{align*}
& \mathrm{max}_{\mathbf z\in\mathcal Z}\,\,\,\mathbf z^\top(H^\top H)^{-1}\mathbf z+\mathbf z^\top\Psi(\mathbf x, H) \mathbf z \\
& =  \mathrm{max}_{\mathbf z\in\mathcal Z}\,\,\, \mathrm{tr}\left( (H^\top H)^{-1}\mathbf z\mathbf z^\top\right)+ \mathbf x ^\top \Upsilon (\mathbf z, H) \mathbf x \\
&\geq n^{-1}\sum^{n}_{i=1}\left(\mathrm{tr}\left( (H^\top H)^{-1}\mathbf h_i\mathbf h^\top_i\right)+ \mathbf x ^\top \Upsilon (\mathbf h_i, H) \mathbf x\right) \\
&= n^{-1}\mathrm{tr}\left( (H^\top H)^{-1}\sum^{n}_{i=1}\mathbf h_i\mathbf h^\top_i\right) \\
&+n^{-1}\mathbf x ^\top\left[ (H(H^\top H)^{-1}H^\top) \circ (H(H^\top H)^{-1}\sum^n_{i=1}\mathbf h_i (\mathbf{h}_i)^\top
(H^\top H)^{-1}H^\top) \right]\mathbf x\\
& = \frac{p}{n}+\frac{1}{n} \mathbf x^\top\left[(H(H^\top H)^{-1}H^\top)\circ (H(H^\top H)^{-1}H^\top)\right]\mathbf x. \qquad \blacksquare
\end{align*}
\endproof
Therefore, we settle to optimize the above lower bound, which depends on $\mathbf x$ only, instead of the inner maximization problem of~\eqref{eq:surrogate}. This results in the following single-level optimization problem of $\mathbf x$:
\begin{align}\label{surrogate-LB}
\mathrm{min}&~\mathbf x^\top\left[(H(H^\top H)^{-1}H^\top)\circ (H(H^\top H)^{-1}H^\top)\right]\mathbf x,\\
         \text{s.t. } &-1 \leq \sum^n_{i=1} x_i\leq 1,\nonumber \\
         &\mathbf x\in \{-1, +1\}^n.\nonumber 
\end{align}

Optimization problem~\eqref{surrogate-LB} is a convex quadratic integer program (matrix $(H(H^\top H)^{-1}H^\top)\circ (H(H^\top H)^{-1}H^\top)$ in the objective of~\eqref{surrogate-LB} is PSD), which can be handled by certain off-the-shelf optimization solvers such as Gurobi.

\section{Numerical Results} \label{se:numerical}
In this section, we present numerical results of our proposed algorithms on standard benchmarks on synthetic and real-world data sets. In order to streamline the exposition, we consider the following labels for our proposed algorithms and the benchmark algorithm:
\begin{itemize}
\item EXACT: the cutting-plane based approach for solving the surrogate model~\eqref{eq:surrogate} described in Section~\ref{sse:exact}. We use a time limit of $300$ seconds as the time limit for this approach.
\item LB\_APPROX: the lower bound approximation for the surrogate model~\eqref{eq:surrogate} described in Section~\ref{sse:LB}.
\item RAND: a standard (re-)randomization technique used in design of experiments. In the randomized allocation approach, we randomly allocate $n/2$ treatments of option $-1$ and $n/2$ treatments of option $1$ to $n$ patients (assuming $n$ is an even number).
\end{itemize}

We evaluate different approaches using the objective value of the original problem in \eqref{eq:var} (we refer to it as the ``Original Objective Value'') and the objective value of the surrogate model in \eqref{eq:surrogate} (we refer to it as the ``Surrogate Objective Value''). For any given allocation $\mathbf x$, the optimal value of the inner problem of the original problem~\eqref{eq:var} is given by:
\[
\mathrm{max}_{\mathbf z\in \mathcal Z}\,\,\,\mathbf z^\top \Sigma_\beta(\mathbf x, H) \mathbf z,
\]
and the optimal value of the inner problem of the surrogate model~\eqref{eq:surrogate} is given by:
\[
\mathrm{max}_{\mathbf z\in \mathcal Z}\,\,\,\mathbf z^\top (H^\top H)^{-1} \mathbf z+\mathbf z^\top\Psi (\mathbf x , H) \mathbf z.
\]
We demonstrate the approximation accuracy of the surrogate objective to the original objective numerically in Appendix \ref{sec:compare}. Considering both objectives allows us to observe whether the performance of each algorithm with respect to the surrogate model is different from the ones with respect to the original objective function.

For RAND, we generate 100 random allocations. The 1\%, 5\%, and 50\% quantiles of the values from the 100 random allocations are denoted by ``RAND(1\%)'', ``RAND(5\%)'', and ``RAND(50\%)'', which are compared with the optimal objective values (including the ``Original Objective Value'' and the ``Surrogate Objective Value'') obtained by EXACT and LB\_APPROX. 

For personalized medicine, it is crucial to investigate the performance of the proposed approaches at the individual level.
Notice that, our optimization problem is formulated to optimize the worst case scenario among all covariates. Therefore, it does not guarantee that the optimal designs can achieve 
better accuracy for each individual. To investigate the performance at the individual level, we compute the variance of $\mathbf z^\top \hat{\pmb\beta}(\mathbf x, H)$ associated with the computed optimal design and compare it with the mean variance associated with random designs for a randomly generated set of patient information $\mathbf z$. Given a patient information vector $\mathbf z_0$, the mean variance of the interaction effect from random designs is 
\begin{equation}\label{eq:var_mean}
\mathrm{E}_{\mathbf x} \left[\mathbf z^\top_0 \Sigma_\beta(\mathbf x, H) \mathbf z_0\right] = \mathbf z^\top_0 \left[\mathrm{E}_{\mathbf x}\Sigma_\beta(\mathbf x, H)\right] \mathbf z_0,
\end{equation}
where the expectation is taken with respect to the random design, and can be approximated empirically, e.g., via a Monte Carlo sample. Given an optimal design, $\mathbf x^\ast$, e.g., computed by approach EXACT or LB\_APPROX, the variance 
of the interaction effect is
\begin{equation}\label{eq:var_opt}
\mathbf z^\top_0 \Sigma_\beta(\mathbf x^\ast, H) \mathbf z_0.
\end{equation}
For a given vector $\mathbf z_0$ of patient information, the percentage of variance reduction yielded by the optimal design $\mathbf x^\ast$ compared to random designs can be expressed by
\begin{equation}\label{eq:var_red}
100\times\frac{\mathbf z^\top_0 \left[\mathrm{E}_{\mathbf x}\Sigma_\beta(\mathbf x, H)\right] \mathbf z_0-\mathbf z^\top_0 \Sigma_\beta(\mathbf x^\ast, H) \mathbf z_0}{\mathbf z^\top_0 \left[\mathrm{E}_{\mathbf x}\Sigma_\beta(\mathbf x, H)\right] \mathbf z_0}\%.
\end{equation}
We randomly generate $1000$ random designs to empirically estimate the above variance reduction measure for 1000 randomly generated patient information vector $\mathbf z_0$'s. Empirically, this variance reduction indicates the extent at which the accuracy of the estimated interaction effect in~\eqref{eq:lm} is improved by the proposed approach compared to a random allocation, which in turn shows the value of the proposed approach in accurately selecting personalized treatment.

In the reminder of this section, we present the numerical performances of each approach on synthetic data sets in Section~\ref{sec:syn} and real-world data sets in Section~\ref{sec:real}, respectively.

\subsection{Numerical Study on Synthetic Data Sets}\label{sec:syn}

We first consider synthetic data sets with randomly generated covariates matrix $H$ in \eqref{eq:target}. Recall that $H$ is an $n\times p$ matrix. The first column of $H$ is loaded by ones, and we randomly generate the entries of the remaining $p$ columns with $-1$ or $1$ of equal probability. 

\begin{figure}
\centering
\includegraphics[scale=0.8]{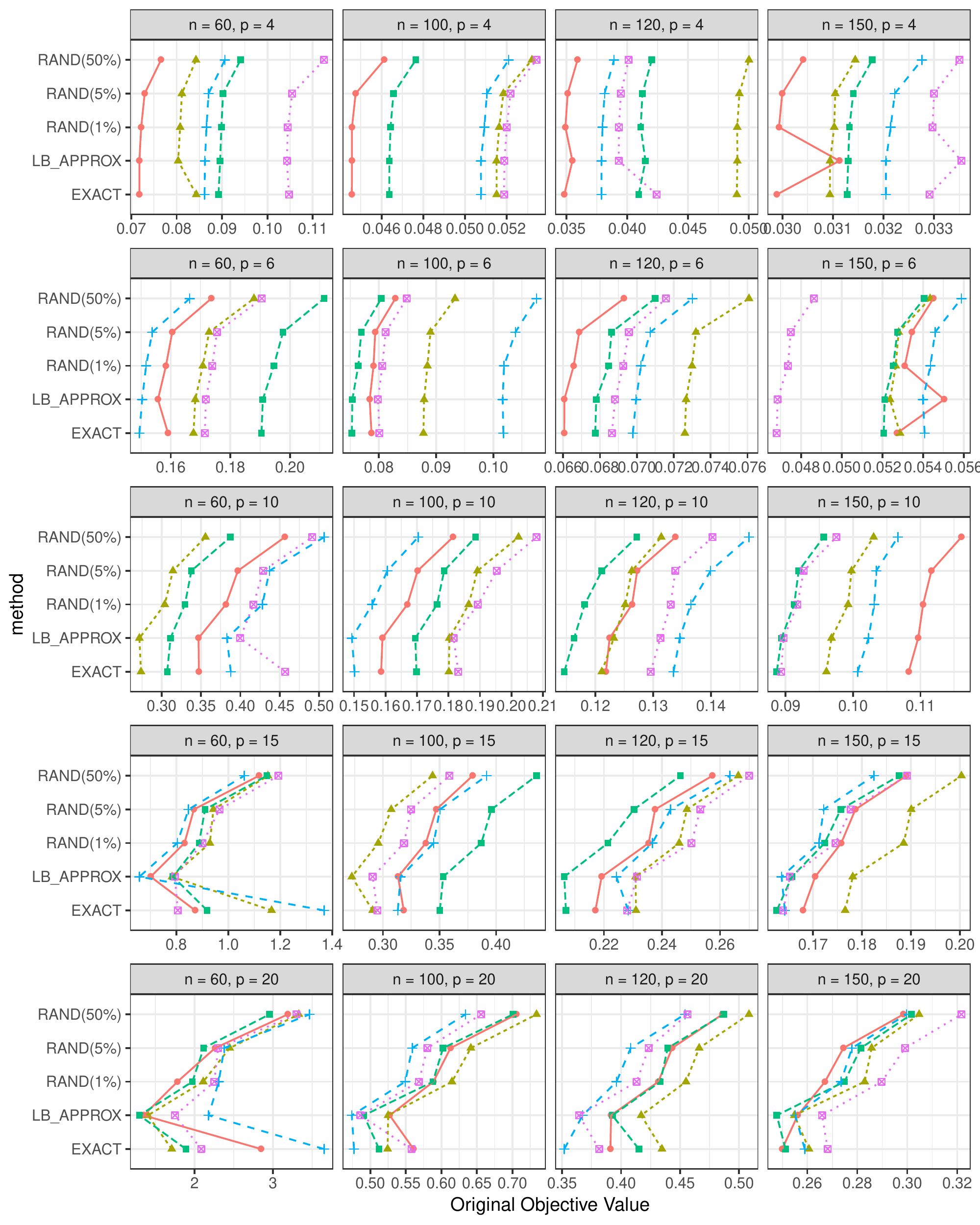}
\caption{The objective values of the original problem with different $n$ and $p$ values. Each color represents the results from one realization of $H$.}\label{fg:true150}
\end{figure}

\begin{figure}
\centering
\includegraphics[scale=0.8]{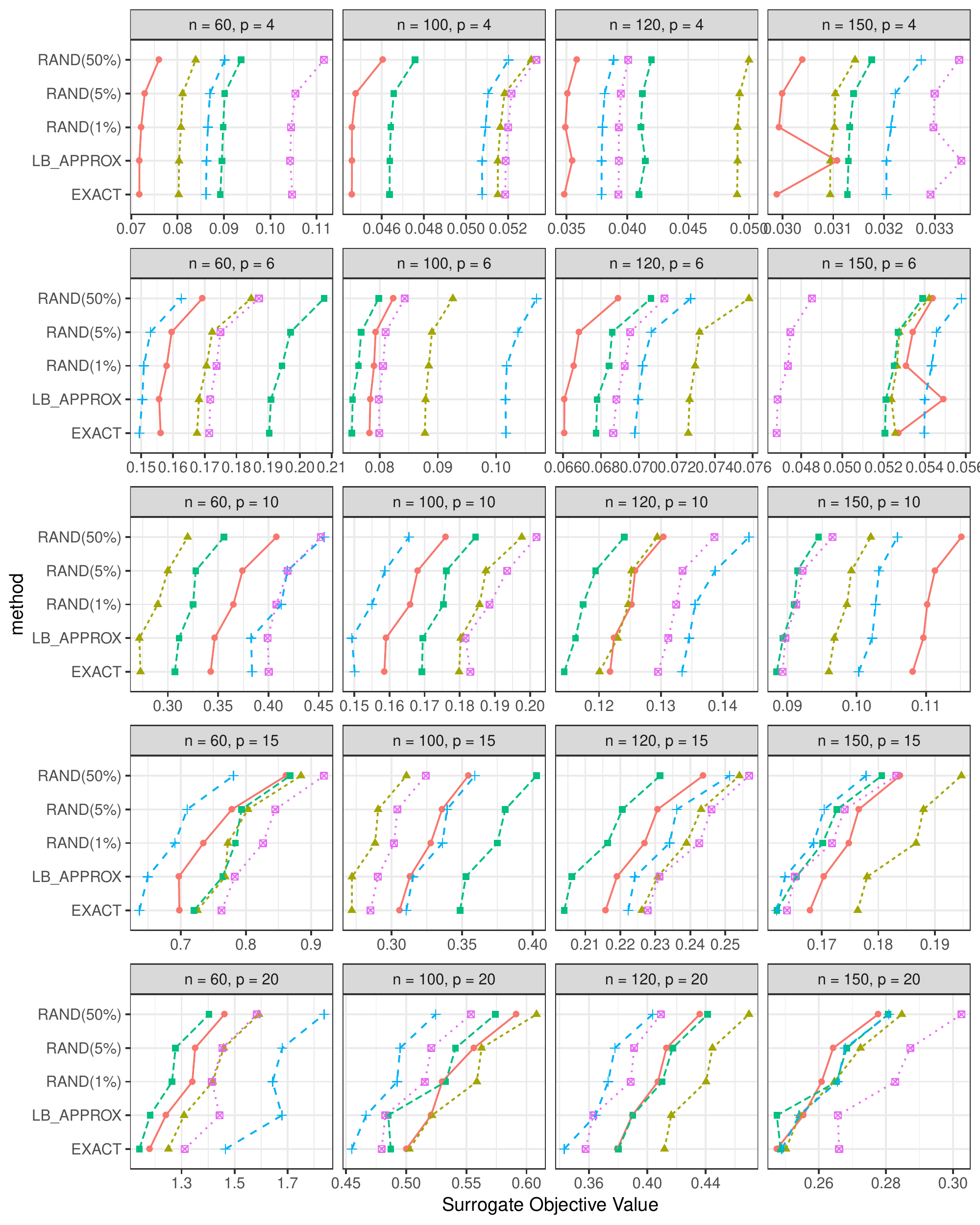}
\caption{The objective values of the surrogate  with different $n$ and $p$ values. Each color represents the results from one realization of $H$.}\label{fg:surrogate150}
\end{figure}

We first consider the performances of different approaches with small $n$ and $p$ values: $n\in \{60, 100, 120, 150\}$, and $p\in \{4, 10, 15, 20\}$. In order to provide a variety of estimates for the objective functions with respect to the random matrix $H$, we consider five randomly generated $H$ matrices for each $n$ and $p$ combination. The  objective values of the original optimization problem \eqref{eq:var} are depicted in Figure~\ref{fg:true150} for different algorithms, and the objective values of the surrogate model are depicted in Figure~\ref{fg:surrogate150}. In these two figures, each color represents results from one realization of the covariates matrix $H$ (recall that we consider five realizations for each combination). From Figures~\ref{fg:true150} and~\ref{fg:surrogate150}, we observe that both EXACT and LB\_APPROX provide more competitive results compared to the random allocation. 
We also observe that although EXACT produces the smallest objective value for the surrogate model in most instances, 
LB\_APPROX is more robust in producing smaller objective values for both the original and the surrogate model. 
A few more observations from these two figures are in order:
\begin{itemize}
    \item There are a few cases, especially when $n$ is relatively small compared to $p$ (e.g., $n=60$ and $p=15$ or $20$), that the solution from EXACT is inferior compared to RAND with respect to the original objective value. The reason is that in these cases, the surrogate model is a poor approximation to the original problem.
    \item When both $n$ and $p$ are large (e.g., $n=120$ or 150 and $p=20$), LB\_APPROX may outperform EXACT in terms of both original and surrogate model objective values. The reason is that EXACT is time-consuming to solve for these instances, e.g., we observe that each iteration of the cutting plane algorithm often exceeds the given time limit of $300$ seconds. Thus, the reported solution from EXACT may be a suboptimal solution of the surrogate model on these instances.    
\end{itemize}

\begin{figure}[t]
\centering
\includegraphics[scale=0.8]{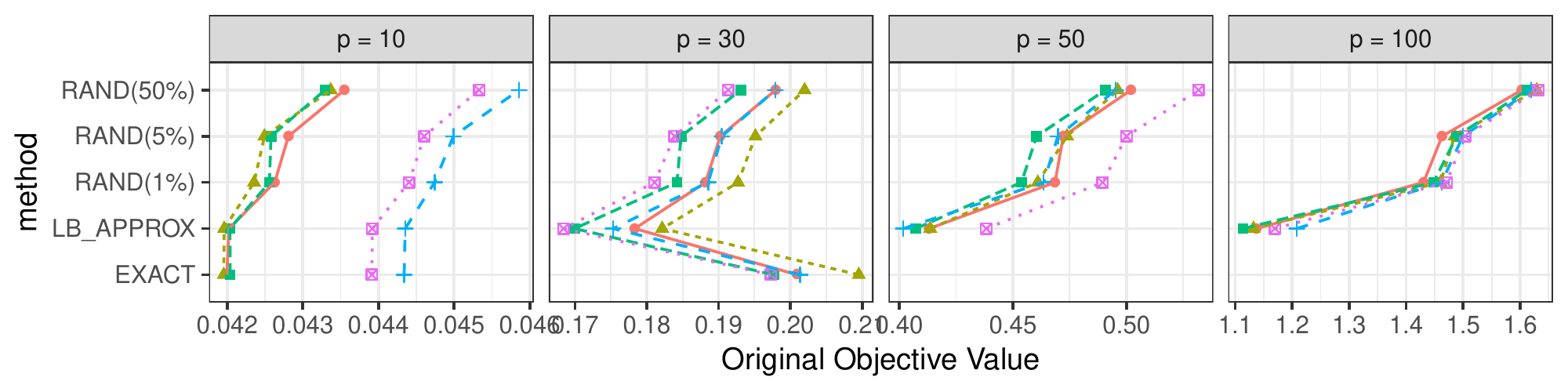}
\caption{The objective values of the original problem with $n=300$. Each color represents the results from one realization of $H$.}\label{fg:n300}
\end{figure}

\begin{figure}[t]
\centering
\includegraphics[scale=0.8]{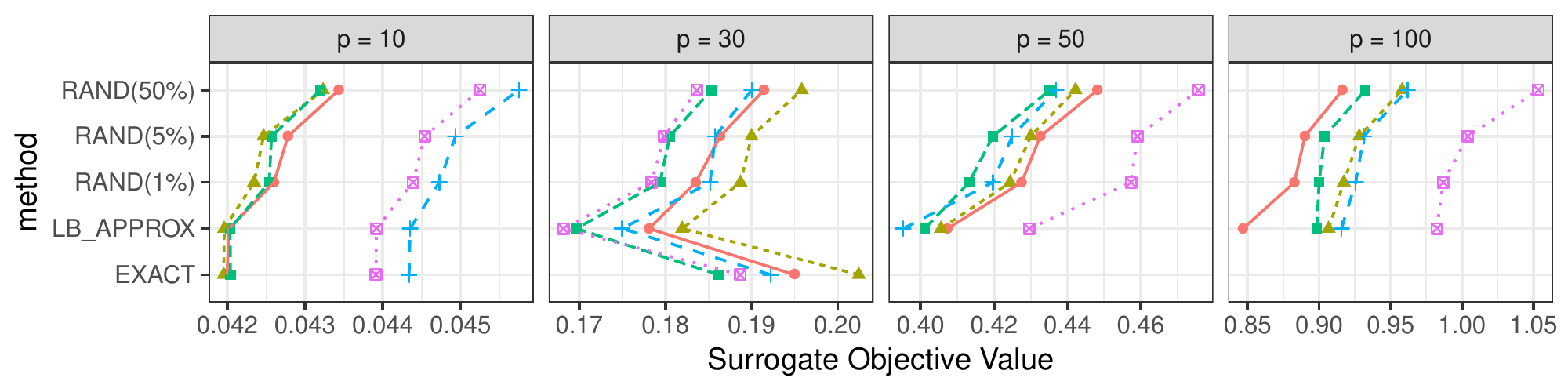}
\caption{The objective values of the surrogate model with $n=300$. Each line represents the results from one realization of $H$.}\label{fg:n300surrogate}
\end{figure}

We now consider large-scale instances with $n=300$. When $n$ is greater than $150$, and $p$ is greater than $30$, the EXACT algorithm becomes computationally intractable. Therefore, we only compare the performance of LB\_APPROX with that of RAND for $p=50$ and $100$. The objective values of the original problem and the surrogate model are depicted in Figures \ref{fg:n300} and \ref{fg:n300surrogate}, respectively. 
As can be seen from Figure~\ref{fg:n300}, for $p=4$ and $p=10$, EXACT and LB\_APPROX outperform randomized algorithms in terms of both the original and surrogate objective values. For $p=30$, EXACT performs poorly with respect to both objectives because only a suboptimal solution of a low quality is available when the time limit is reached. On the other hand, LB\_APPROX outperforms randomized algorithms and produces robust performance in all ranges of $p$ with respect to both original and surrogate models.

\begin{figure}[t]
\centering
\includegraphics[scale=0.65]{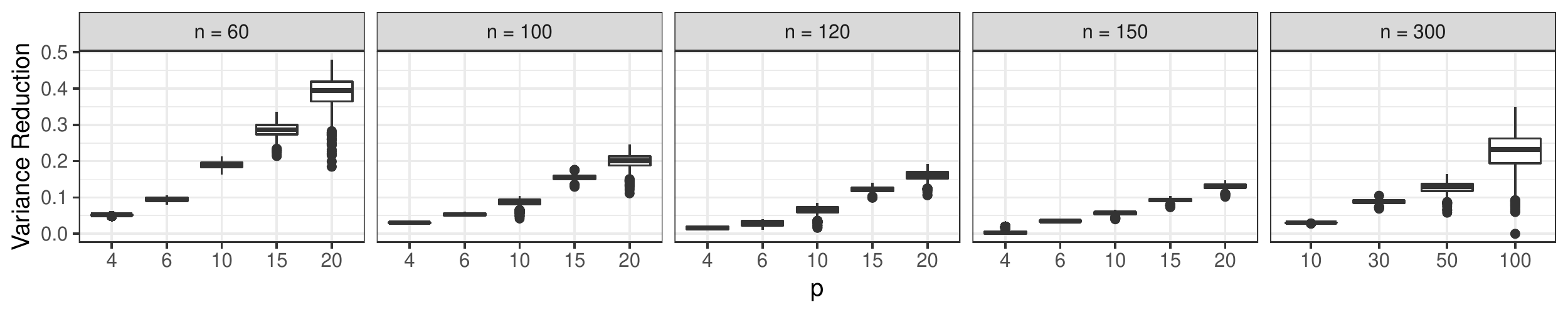}
\caption{The percentage of variance reduction of the design attained from LB\_APPROX with respect to the mean of the random designs.}\label{fg:var_LP}
\end{figure}

\begin{figure}[t]
\centering
\includegraphics[scale=0.65]{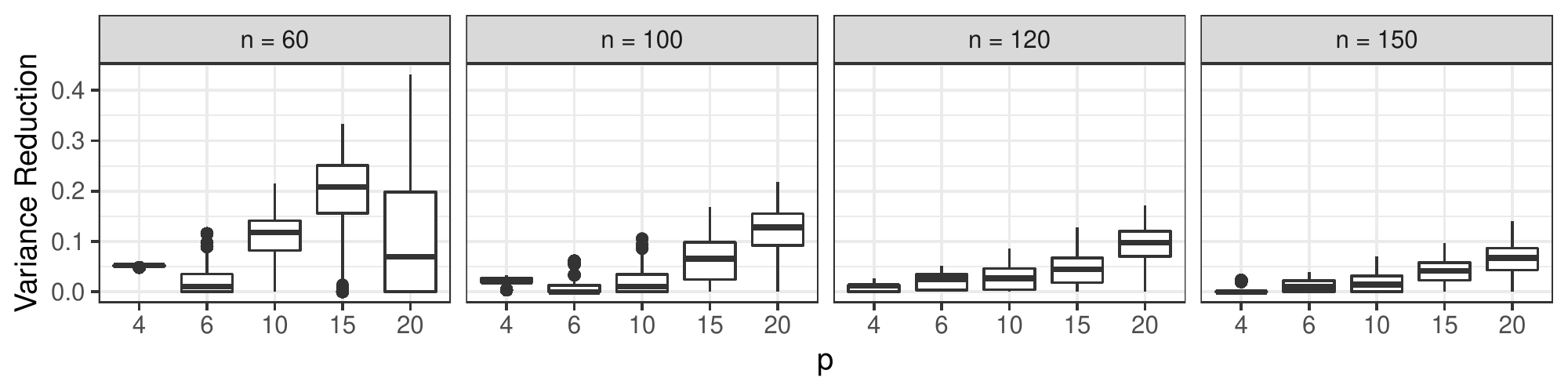}
\caption{The percentage of variance reduction of the design attained from EXACT with respect to the mean of the random designs.}\label{fg:var_exact}
\end{figure}

We now evaluate the variance reduction measure in \eqref{eq:var_red} of the LB\_APPROX with respect to the mean variance from random designs. By comparing \eqref{eq:var_mean} and \eqref{eq:var_opt} over 1000 randomly generated patient instances $\mathbf z_0$, almost 100\% patients achieve smaller variance of the estimated interaction effect by using the optimal design from LB\_APPROX. As shown in Figure \ref{fg:var_LP}, the percentage of variance reduction ranges from 5\% to 50\% for different individuals. The variance reduction results from EXACT are shown in Figure \ref{fg:var_exact}, the percentage of variance reduction ranges from 1\% to 40\% for different individuals. Over different cases, 50\%--90\% patients achieve smaller variance of the estimated interaction effect by using the proposed optimal design compared to random design.

\subsection{Case Study}\label{sec:real}
Warfarin is an anticoagulant medication, which is used to treat blood clots. In the US, more than 30 million patients were prescribed Warfarin in 2010 \citep{ClinCalc2016}. However, taking an incorrect dose of Warfarin can cause significant adverse effects \citep{wysowski2007bleeding}. Therefore, there has been significant interest from the medical community to improve dose prescription strategies by taking patients covariates into account. In particular, the International Warfarin Pharmacogenetics Consortium collected clinical and genetic data from 5700 patients who were treated with Warfarin \citep{international2009estimation}. This data set was used to design a personalized dosing algorithm and it is publicly available. Their analysis shows that the following covariates are significant: age, height, weight, race, use of enzyme inducers, use of amiodarone, VKORC1, and CYP2C9. Specifically, the VKORC1 gene provides the instructions to produce an enzyme that activates clotting proteins, and the CYP2C9 gene provides the instructions to produce an enzyme that helps protein processing. This result confirms that genetic factors can play a notable role in optimal Warfarin dosage \citep{white2010patient}.

In our case study, we consider the optimal design of the aforementioned trial retrospectively. That is, if the decision makers were to design the dose-finding trial with the covariates that they observed, what would have been the optimal way? Note that in the case study there were three dosages: low ($\le$ 21 mg per week), medium ($>$ 21 and $<$49 mg per week), and high ($\ge$ 49 mg per week). Because we only consider two levels, we extract the data for patients that had low and high prescription. We assume that the covariates are those that are considered significant in the literature and mentioned above. Following the results given by~\cite{international2009estimation}, age is categorized to nine groups ([10,20), [20,30),..., [90,-)), height to three groups ([0, 160), [160, 180), [180, -)), weight to three groups ([0,60), [60, 90), [90, -)), race to four groups (White, Asian, Black, and others), use of enzyme inducer is binary (Yes, No), use of amiodarone is binary (Yes, No), VKORC1 to three groups (A/A, A/G, G/G), and CYP2C9 to six groups ($^*1/^*1$, $^*1/^*2$, $^*1/^*3$, $^*2/^*2$, $^*2/^*3$, $^*3/^*3$). By excluding the patients with missing/censored data, we have the data for 1476 patients and 21 covariates.

For this large data set, EXACT algorithm is computationally intractable, therefore, we only compare the performances of LB\_APPROX with RAND.
The results are provided in Table \ref{tb:all}.  
We observe that LB\_APPROX outperforms RAND in terms of both the original and the surrogate model objective value. This observation is consistent with the results shown in Section \ref{sec:syn}. Over 1000 randomly generated patient information $\mathbf z_0$, LB\_APPROX achieves variance reduction for all 1000 patients, and the percentage of variance reduction of LB\_APPROX ranges from 1\% to 8\%. This modest variance reduction is somewhat expected, since when $n$ is much larger than $p$, as is the case here, random design can already perform well in fitting the linear model~\eqref{eq:lm}, which does not leave much room for further improvement.

\begin{table}[ht]
\centering
\caption{The objective values of the real data set with 1476 patients.}\label{tb:all}
\begin{tabular}{rrr}
  \hline
Method & Original Objective Value& Surrogate Objective Value\\ 
  \hline
LB\_APPROX & 0.8265 (0\%) & 0.8265 (0\%)\\ 
RAND(1\%) & 0.8316 & 0.8314 \\ 
RAND(5\%) & 0.8340 & 0.8337 \\ 
RAND(50\%) & 0.8522 & 0.8502 \\ 
   \hline
\end{tabular}
\end{table}

To evaluate the performance of EXACT and compare it with alternative approaches on this real data set, we truncate the problem size by randomly selecting 100 patients to conduct a relatively small scale experiment. To ensure that there is no numerical issue in evaluating the true objective, we only include 17 columns of covariates matrix out of 21 in this experiments. The results are given in Table \ref{tb:small}. The results show that
EXACT algorithm outperforms other algorithms in terms of both the original model and the surrogate model objective value. The LB\_APPROX algorithm is located at or under the 1\% quantile among the objective values generated from 100 random designs.  

\begin{figure}[t]
\centering
\includegraphics[scale=0.8]{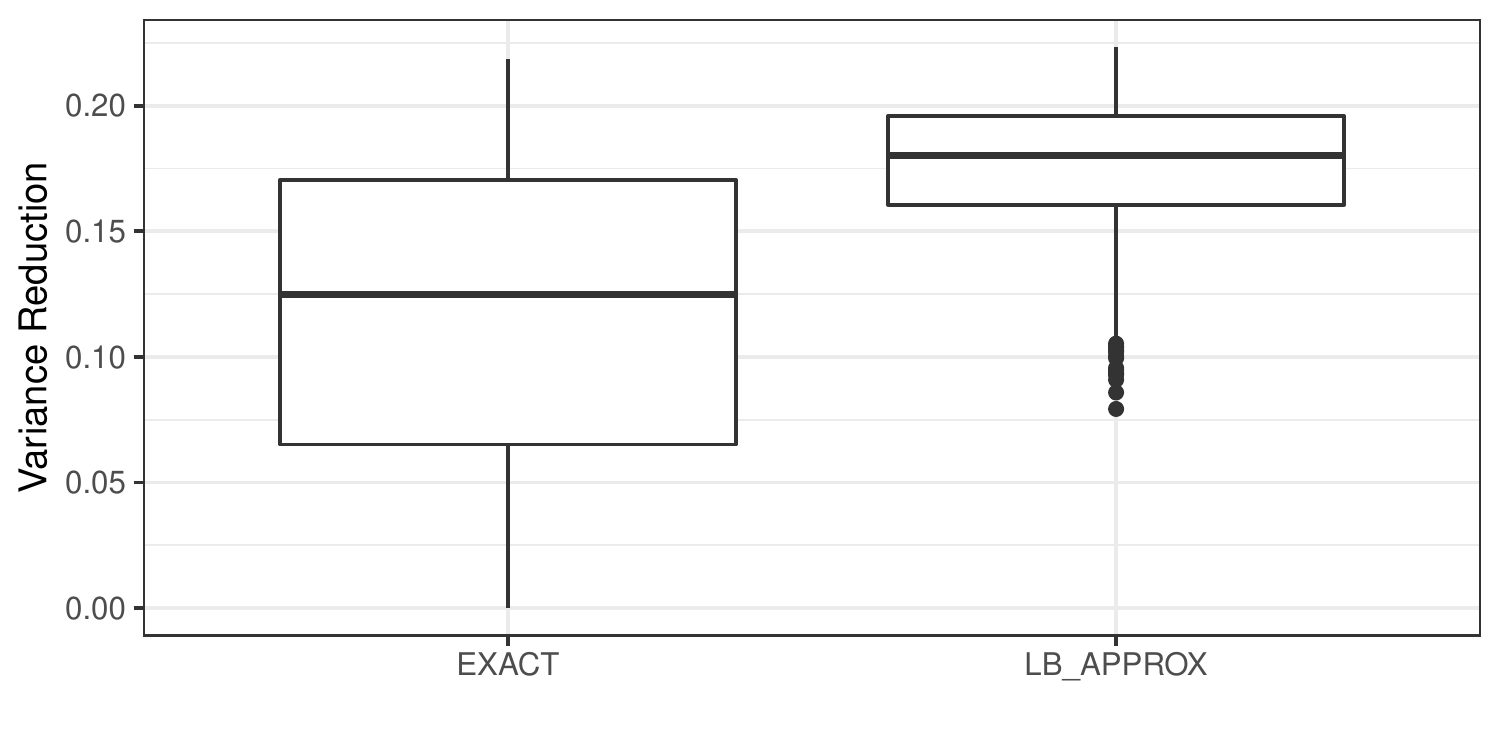}
\caption{The percentage of variance reduction of the design attained from LB\_APPROX and EXACT with respect to the mean of the random designs for the real example with $n=100$.}\label{fg:real}
\end{figure}

Over 1000 randomly generated patient information $\mathbf z_0$, LB\_APPROX reduces the variance of all 1000 patients with respect to the mean variance, whereas EXACT reduces the variance of 896 patients with respect to the mean variance. 
We show the boxplot of the percentage of variance reduction for both LB\_APPROX and EXACT in Figure \ref{fg:real}.

\begin{table}[h]
\centering
\caption{The objective values of the real data set with 100 patients.}\label{tb:small}
\begin{tabular}{rrr}
  \hline
Method & Original Objective Value& Surrogate Objective Value\\ 
  \hline
  EXACT & 6.9999 (0\%) & 6.8642  (0\%)  \\ 
LB\_APPROX & 7.1715  (0\%)  & 7.1396   (1\%) \\ 
RAND(1\%) & 7.5634 & 7.2928 \\
RAND(5\%) & 7.8390 & 7.4735 \\
RAND(50\%) & 9.2347 & 8.2819 \\  \hline
\end{tabular}
\end{table}

\section{Conclusion} \label{se:conclusion}
This study introduced a novel model to incorporate patient covariates into treatment effect as significant evidence is established in precision medicine literature that patients may respond differently to a treatment. We studied the optimal design of two-armed clinical trials using the introduced model, which help practitioners design clinical trials that more accurately estimate treatment effects. Our extended model posed significant challenges in the optimization problems emanated from optimal design of such experiments, which has optimization over design and patient covariates simultaneously. In particular, we minimized (over design) the maximum (over patient covariates) variance of treatment effect, which is a min-max bi-level mixed-integer nonlinear program. We proposed a solution methodology by replacing the variance of treatment effect with its natural approximation, motivated by asymptotically balanced trials. We proposed an exact algorithm to solve the surrogate optimization problem via reformulation and decomposition techniques. In addition, we created a lower bound for the inner optimization problem and solved the outer optimization over the lower bound. We tested our algorithms on hypothetical and real-world data sets. Our numerical analysis concluded the following insights: 1- The quality of approximation for the surrogate objective function is high if the number of covariates is small compared to the total number of patients in the trial, that is, low-dimensional settings. 2- Our proposed algorithms outperform the standard (re-)randomization techniques used in the optimal design literature. Our result echoes that of \citet{bertsimas2015power}, which showed the power of optimization over randomization in a different optimal design setting. 3- Our lower bounding algorithm produced high-quality solutions with respect to the surrogate and original objective function across all settings. This observation suggests that the lower bounding algorithm, which is easy to implement via off-the-shelf solvers, can be used instead of the proposed exact algorithm in practice. 

Finally, our modeling approach to incorporate patient covariates into treatment effect can be generalized to other settings. In particular, our framework generalizes the A-B testing framework in \citet{bhat2017near} in incorporating side information into treatment effect. Therefore, it can be used for other applications such as e-commerce, on-line advertising, and assortment, where one seeks to investigate covariate-dependent treatment effect. Furthermore, our framework is flexible in incorporating a variety of operational constraints. Specifically, one can add constraints on design variables $\mathbf x$ into the outer optimization problem and all of our methodology still holds. This is an appealing feature because practitioners may face some limitations in the design phase upfront and our methodology can provide robust solutions in those settings.

 \appendix

 \section{Numerical comparison between the original objective value and the surrogate objective value}\label{sec:compare}

\begin{figure}
\centering
\includegraphics[scale=0.8]{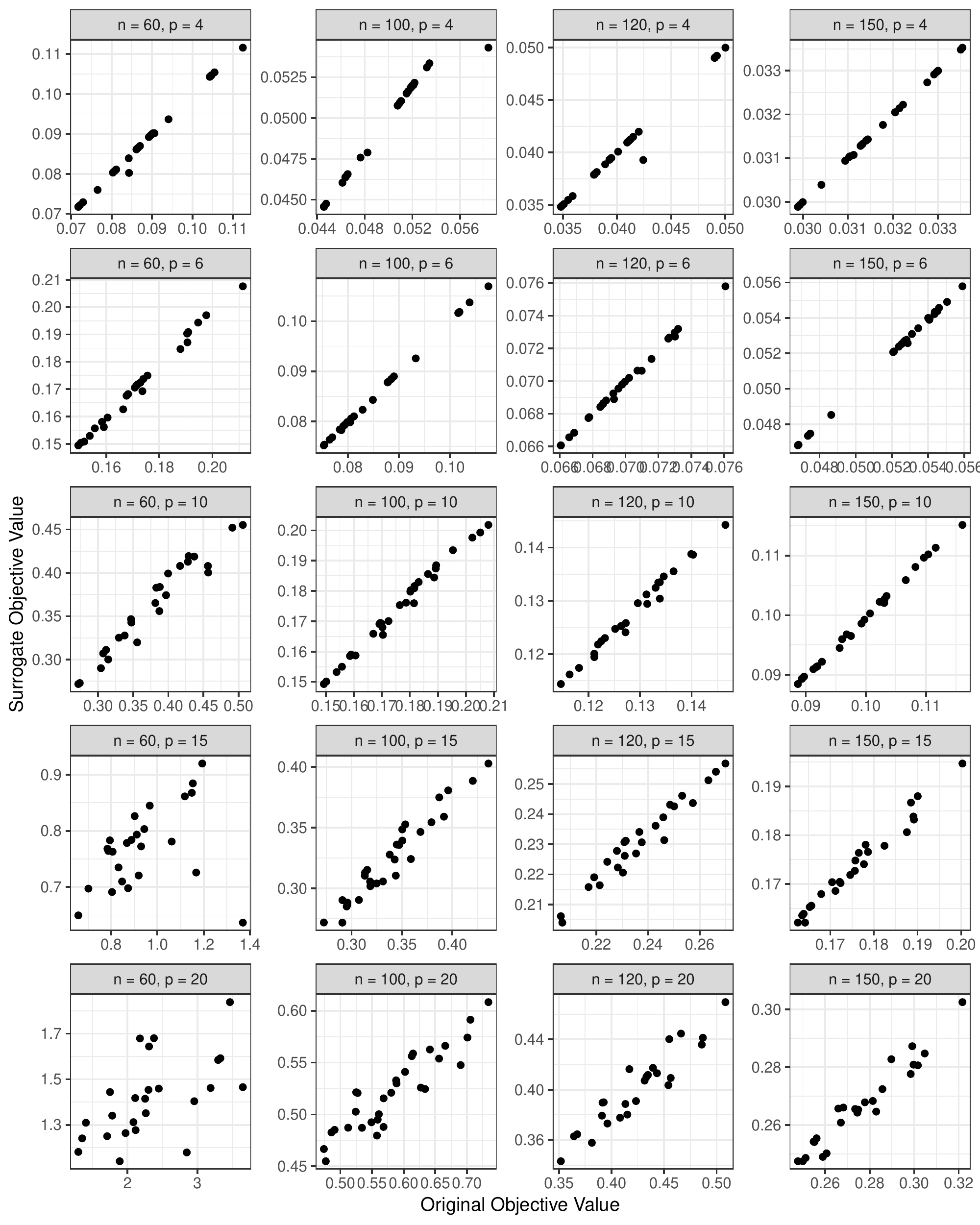}
\caption{Original objective value verse surrogate objective value for the instances in Figure \ref{fg:true150}.}\label{fg:compare_small}
\end{figure}
 
\begin{figure}
\centering
\includegraphics[scale=0.8]{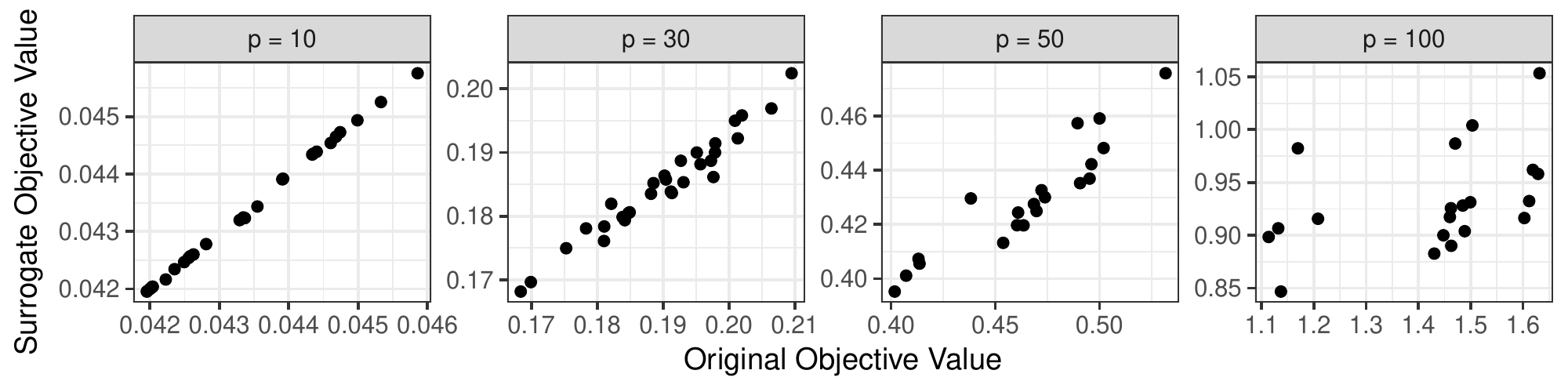}
\caption{Original objective value verse surrogate objective value for the instances in Figure \ref{fg:n300}.}\label{fg:compare_large}
\end{figure}

We provide a numerical comparison between the original objective value and the surrogate objective value in this section. 
Figures \ref{fg:compare_small} and \ref{fg:compare_large} show the results of the instances as in Figures \ref{fg:true150} and \ref{fg:n300}, respectively. 
In each subfigure, the x-axis is the value of the original objective, and the y-axis is the value of the surrogate objective. The ideal case to make a good approximation
from the surrogate model to the original problem is that the dots are roughly aligned along the diagonals. The results in  Figures \ref{fg:compare_small} and \ref{fg:compare_large}  show that, the surrogate model is a good approximation to the original problem when $p$ is relatively small compared to $n$, whereas
the approximation is not accurate if $p$ is relatively large compared to $n$.

\end{document}